\documentclass[10pt,twoside]{Lowx2011}
\usepackage{epsf,amsmath}
\usepackage{graphicx}
\usepackage{mcite}
\usepackage[utf8x]{inputenc}
\usepackage[english]{babel}
\usepackage{bbm}
\if@mathematic
   \def\vec#1{\ensuremath{\mathchoice
                     {\mbox{\boldmath$\displaystyle\mathbf{#1}$}}
                     {\mbox{\boldmath$\textstyle\mathbf{#1}$}}
                     {\mbox{\boldmath$\scriptstyle\mathbf{#1}$}}
                     {\mbox{\boldmath$\scriptscriptstyle\mathbf{#1}$}}}}
\else
   \def\vec#1{\ensuremath{\mathchoice
                     {\mbox{\boldmath$\displaystyle#1$}}
                     {\mbox{\boldmath$\textstyle#1$}}
                     {\mbox{\boldmath$\scriptstyle#1$}}
                     {\mbox{\boldmath$\scriptscriptstyle#1$}}}}
\fi

\begin{document}

\title{High-Energy Dijets in ${\cal N}=4$ SYM
\thanks{Research supported by E. Comission (LHCPhenoNet (PITN-GA-2010-264564)) and C. Madrid (HEPHACOS ESP-1473).}}

\author{ G. CHACHAMIS$^{\star}$, \underline{J.D. MADRIGAL}$^\dagger$ and A. SABIO VERA$^{\dagger}$\\ \\
$^{\star}$Paul Scherrer Institut, CH-5232 Villigen PSI, Switzerland.\\
$^\dagger$Instituto de Física Teórica UAM/CSIC and\\
Universidad Autónoma de Madrid, E-28049 Madrid, Spain.\\
}

\maketitle

\begin{abstract}
\noindent Ratios of azimuthal angle correlations for Mueller-Navelet jets are compared in QCD and ${\cal N}=4$ SYM. Such observables are well suited to study the effects of Möbius invariance in the Regge limit. The role of the renormalization prescription  and the size of conformal contributions is addressed, showing that the BLM procedure best reproduces the ${\cal N}=4$ SYM results.
\end{abstract}



\markboth{\large \sl \hspace*{0.25cm}Chachamis, Madrigal and Sabio Vera
\hspace*{0.25cm} Low-$x$ Meeting 2011} {\large \sl \hspace*{0.25cm} High-Energy Dijets in ${\cal N}=4$ SYM}

\section{Introduction}

The idea that some lessons can be drawn about collider physics and QCD from its maximally supersymmetric version ${\cal N}=4$ Super Yang-Mills (MSYM) has been extensively studied in the last years (see e.g. \cite{HM2008,HIM2008,Hat2008,LMKS2009,BKVZ2009}). In fact the large amount of symmetry of MSYM translates into amazing simplicity of its scattering amplitudes \cite{MP1991,Wit2004,BDS2005}. Though it is neither conformal nor supersymmetric, ${\cal N}=4$ SYM is very similar to QCD in most respects. MSYM contributions to QCD amplitudes turn out to give the pieces with highest ``degree of transcendentality" \cite{KL2003}. Moreover, in the Regge limit, both theories are identical in leading logarithmic approximation. Therefore, the study of observables allowing a sensible comparison of QCD and ${\cal N}=4$ SYM, especially in the high-energy limit, is of great interest to quantify how much one can trust ${\cal N}=4$ results ---in particular, those obtained through AdS/CFT duality \cite{Mal1998}--- describing collider physics.

\section{Azimuthal Correlations in QCD and ${\cal N}=4$ SYM}

The observables considered in \cite{ACMS2011} to compare QCD and ${\cal N}=4$ SYM in the Regge limit are ratios of azimuthal correlations between two forward jets with similar transverse momenta $\vec{p}_{1,2}^2$ produced at large rapidity separation $Y\sim\ln(x_1x_2s/\sqrt{\vec{p}_1^2\vec{p}_2^2)},\,s\gg\sqrt{\vec{p}_1^2\vec{p}_2^2)}$ (the so-called Mueller-Navelet jets \cite{MN1987}), with $x_{1,2}$ the fractions of longitudinal momenta of the parent hadrons carried by the jets\footnote{To leading order radiated gluons can be directly identified with jets. The present computation is carried to next-to-leading order (NLO) for the Green's function, while keeping for simplicity jet vertices to leading order ($\Phi_{\rm jet}\simeq\Phi_{\rm jet}^{(0)}$). NLO jet vertices for Mueller-Navelet setup have been included in \cite{CSSW2010}.}. Such ratios were previously studied in QCD \cite{Ver2006,VS2007} where they were shown to exhibit an excellent perturbative convergence and to be essentially independent of parton densities for large rapidity separation, what allows us to compute at partonic level, giving a sound comparison.\\

The BFKL formalism, in which terms of the form\footnote{$\lambda$ denotes the 't Hooft coupling in MSYM and $\bar{\alpha}_s\equiv\alpha_sN_c/\pi$ in QCD.} ${\cal O}((\lambda Y)^n)$ are resummed to all orders in the multi-Regge limit \cite{Lip1976,FKL1975,FKL1976,FKL1977,BL1978}, is best suited to compute the cross-section for Mueller-Navelet jets. At the partonic level, it can be written as a convolution with jet vertices $\Phi_{\rm jet_i}(\vec{q},\vec{p}_i)\simeq\Phi^{(0)}_{\rm jet_i}(\vec{q},\vec{p}_i)=\Theta(\vec{q}^2-\vec{p}_i^2)$, $\vec{p}_i^2$ being a resolution scale

\begin{equation}
\label{cross}
\hat{\sigma}(\alpha_s, Y, p_{1,2}^2) =
\int d^2\vec{q}_1\int d^2\vec{q}_2 ~\Phi_{\rm jet_1} (\vec{q}_1, p_1^2)\frac{d\hat{\sigma}}{d^2\vec{q}_1 d^2\vec{q}_2}\Phi_{\rm jet_2} (\vec{q}_2, p_1^2),
\end{equation}

and the differential cross section is simply given in terms of the Mellin transform $f(\vec{q}_1,\vec{q}_2,Y)=\int\frac{d\omega}{2\pi i}e^{\omega Y}\tilde{f}(\vec{q}_1,\vec{q}_2,\omega)$ of the solution to the BFKL equation $\omega\tilde{f}(\vec{q}_1,\vec{q}_2,\omega)=\delta^{(2)}(\vec{q}_1-\vec{q}_2)+\int d^2\,\vec{\kappa}\,{\cal K}(\vec{q}_1,\vec{\kappa})\tilde{f}(\vec{\kappa},\vec{q}_2,\omega)$ by 

\begin{equation}
\frac{d\hat{\sigma}}{d^2\vec{q}_1 d^2\vec{q}_2}=\frac{\pi^2\bar{\alpha}_s^2}{2}\frac{f(\vec{q}_1,\vec{q}_2,Y)}{\vec{q}_1^2\vec{q}_2^2}.
\end{equation}

The kernel (at NLO) is, in the basis of normalized LO eigenfunctions $\langle\vec{q}|n,\nu\rangle=\frac{1}{2\sqrt{\pi}}(\vec{q}^2)^{i\nu-1/2}e^{in\vartheta}$ (with LO eigenvalues $\chi_0(n,\nu)=2\psi(1)-\psi\left(\frac{1+n}{2}+i\nu\right)-\psi\left(\frac{1+n}{2}-i \nu\right)$)

\begin{equation}
\begin{aligned}
\label{kernelnlo}
&\langle n,\nu|\hat{\cal K}|\nu',n'\rangle= \bar{\alpha}_{s, \overline{\rm MS}}\bigg[\chi_0\left(|n'|,\frac{1}{2}+i\nu'\right)\left(1-\frac{\bar{\alpha}_{s, \overline{\rm MS}}\beta_0}{8N_c}\left(i\frac{\partial}{\partial\nu}-i\frac{\partial}{\partial\nu'}-2\ln\mu^2\right)\right)\\
&+\bar{\alpha}_{s, \overline{\rm MS}}\chi_1\left(|n'|,\frac{1}{2}+i\nu'\right)
-\frac{\bar{\alpha}_{s, \overline{\rm MS}}\beta_0}{8N_c}\chi_0\left(|n'|,\frac{1}{2}+i\nu'\right)\bigg]\delta^n_{n'}\delta(\nu-\nu'),
\end{aligned}
\end{equation}

in the QCD case \cite{FL1998,CC1998,KL2000}, while in ${\cal N}=4$ absence of running leads to \cite{KL2000}
\begin{equation}
\langle n,\nu|\hat{\cal K}_{\rm MSYM}|\nu',n'\rangle=\lambda\left[\chi_0\left(|n'|,\frac{1}{2}+i\nu'\right)+\lambda\chi_1^{\rm MSYM}\left(|n'|,\frac{1}{2}+i\nu'\right)\right]\delta^n_{n'}\delta(\nu-\nu'),
\end{equation}

where ($\nu=i(1/2-\gamma)$, ${\cal S}=\left(4-\pi^2+\frac{5\beta_0}{N_c}\right)$ and the function $\Omega$ is given in \cite{KL2000})
\begin{equation}
\begin{aligned}
&\chi_1(n,\gamma)={\cal S}\frac{\chi_0(n,\gamma)}{12}+\frac{3}{2}\zeta (3)-\frac{\beta_0}{8N_c}\chi_0^2(n,\gamma)+\Omega(n,\gamma)-\frac{\pi^2\cos(\pi\gamma)}{4\sin^2(\pi\gamma)(1-2\gamma)}\\
\times & \left[\left(3+\left(1+\frac{N_f}{N_c^3}\right)\frac{2+3\gamma(1-\gamma)}{(3-2\gamma)(1+2\gamma)}\right)\delta_0^n -\left(1+\frac{N_f}{N_c^3}\right)\frac{\gamma (1-\gamma)}{2(3-2\gamma)(1+2\gamma)}\delta^n_2\right];\\
&\chi_1^{\rm MSYM}(n,\gamma)=\frac{1-\zeta (2)}{12}\chi_0(|n|,\gamma)+\frac{3}{2}\zeta (3)+\Omega (|n|,\gamma).
\end{aligned}
\end{equation}

Now, it was shown in \cite{Ver2006,VS2007} that the differential cross section in azimuthal angle
$\phi=\vartheta_1-\vartheta_2-\pi$ ($\vartheta_i$ is the azimuthal angle of each jet), for $\vec{p}^2\simeq\vec{p}_1^2\simeq\vec{p}_2^2$, can be written as

\begin{equation}\label{Cn}
\begin{aligned}
\frac{d\hat{\sigma}(\bar{\alpha}_s, Y, p_{1,2}^2)}{d\phi}&=\frac{\pi^2\bar{\alpha}_s^2}{4\sqrt{p_1^2p_2^2}}\sum_{n=-\infty}^\infty e^{in\phi}\mathcal{C}_n(Y),\\
{\cal C}_n \left(Y\right) &=
\int_{-\infty}^\infty \frac{d \nu}{2 \pi}\frac{e^{{\bar \alpha}_s \left(p^2\right) Y \left(\chi_0\left(\left|n\right|,\nu\right)+{\bar \alpha}_s  \left(p^2\right) \left(\chi_1\left(\left|n\right|,\nu\right)-\frac{\beta_0}{8 N_c} \frac{\chi_0\left(\left|n\right|,\nu\right)}{\left(\frac{1}{4}+\nu^2\right)}\right)\right)}}{\left(\frac{1}{4}+\nu^2\right)}
\end{aligned}.
\end{equation}
The analogous expression for ${\cal C}_n$ in ${\cal N}=4$ SYM is obtained with obvious changes ($\bar{\alpha}_s\to\lambda,\chi_1\to\chi_1^{\rm MSYM}$) and recalling that $\beta_0=0$. In the Fourier decomposition \eqref{Cn}, $n$ is the conformal spin that labels a representation of SL(2,$\mathbbm{C}$). This is the conformal group in two dimensions, and the LO BFKL equation is invariant under it in the transverse plane \cite{Lip1986}. The origin of this symmetry is unclear. Observables related to higher conformal spins, sensitive only to this transverse plane, can probe this SL(2,$\mathbbm{C}$) invariance and, moreover are not affected by the collinear instabilities typical of the $n=0$ component.\\

It is remarkable that such observables can be directly obtained from the coefficients ${\cal C}_n$. For $n=0$, we have the total cross section: $\hat{\sigma}(\vec{p}_{1,2}^2,Y)=\frac{\pi^3\bar{\alpha}_s^2}{2\sqrt{\vec{p}_1^2\vec{p}_2^2}}{\cal C}_0(Y)$. Contributions from higher conformal spins are projected in the correlations
\begin{equation}
\langle\cos (m\phi)\rangle=\frac{{\cal C}_m (Y)}{{\cal C}_0(Y)},\quad{\cal R}_{m,n}(Y)\equiv \frac{\langle\cos (m\phi)\rangle}{\langle\cos (n\phi)\rangle}=\frac{{\cal C}_m (Y)}{{\cal C}_n (Y)}.
\end{equation}

The ratios ${\cal R}_{m,n}$ are introduced to cancel the contribution with $n=0$, so that we can expect them to have an extremely good perturbative convergence. 

\section{The BLM Procedure}

At NLO level, the choice of the renormalization prescription is very important in the comparison between QCD and MSYM. Brodsky, Lepage and Mackenzie (BLM) developed a prescription to set the scale which can be argued to be very  natural for many observables \cite{BLM1983} (see also \cite{BdG2011}). At NLO, a finite renormalization is equivalent to a redefinition of the coupling (e.g. for transition from $\overline{\rm MS}$ to MOM scheme \cite{CG1979}, $\alpha_{\rm MOM}=\alpha_{\overline{\rm MS}}(1+T_{\rm MOM}\,\alpha_{\overline{\rm MS}}/\pi)$) and this in turn to a rescaling of the point at which the coupling is evaluated $\mu\to\bar{\mu}=\mu\exp (-T_{\rm MOM}/2\beta_0)$. In the BLM procedure the scale is set in such a way that the coupling redefinition absorbs all charge renormalization corrections, leaving a perturbative series identical to that of the conformally invariant theory with $\beta=0$. To enhance the effect of BLM in gluon dominated processes, it is appropriate to use a physical scheme for nonabelian interactions. Such an strategy was followed in \cite{BFK+1999}, where BLM was applied to the pomeron intercept in $\gamma^*\gamma^*$, obtaining a result much closer to that expected from phenomenology and hardly sensitive to the transverse scale, approaching conformal behaviour (\textsc{Fig.} \ref{1}). Here MOM scheme was chosen\footnote{We also adopt the MOM scheme for the BLM procedure, with Yennie gauge ($\xi=3)$. To avoid the dependence on $\xi$ it is also possible to use for instance the $\Upsilon$ scheme based on $\Upsilon\to ggg$ decay.}, for which

\begin{equation}
\begin{aligned}
T_{\rm MOM}&=T_{\rm MOM}^{\rm conf}+T_{\rm MOM}^\beta,\quad T_{\rm MOM}^{\rm conf}=\frac{N_c}{8}\left(\frac{17I}{2}+\frac{3\xi}{2}(I-1)+\xi^2\left(1-\frac{I}{3}\right)-\frac{\xi^3}{6}\right),\\
T_{\rm MOM}^\beta &=-\frac{\beta_0}{2}\left(1+\frac{2I}{3}\right),\quad I\simeq 2.3439.
\end{aligned}
\end{equation}

In \cite{ACMS2011} the same procedure is applied to the ratios ${\cal R}_{m,n}$ (see \cite{ACMS2011} for the technical details). We just want to remark that for general conformal spin the value of the BLM scale is

\begin{equation}
\vec{q}^{\rm 2~MOM}_{~~\rm BLM}(n,\nu)=\vec{q}^2\exp\left(\frac{1}{2}\chi_0(n,\nu)+\frac{1+4I}{3}\right)
\end{equation}

The BLM procedure, which produces a high scale for $n=0$ ($\varrho\equiv\vec{q}^{\rm 2~MOM}_{~~\rm BLM}(n,\nu=0)/\vec{q}^2$ is $\sim 105$ for $n=0$), gives a more natural scale for higher conformal spins, e.g. $\varrho(n=1)\sim 33,~\varrho(n=2)\sim 18$.\\

It was expected that conformal contributions, resummed to all orders, would be of great importance. In fact, NLO corrections for the truly conformal ${\cal N}=4$ SYM kernel are approximately only a third of those in QCD. So BLM prescription is expected to make QCD results closer to those of MSYM. If Möbius invariance is related to the 4d conformal symmetry of MSYM this should be clearly seen for our observables.

\section{MSYM versus QCD}

\begin{figure}[h!]
\centering
\includegraphics[scale=0.8]{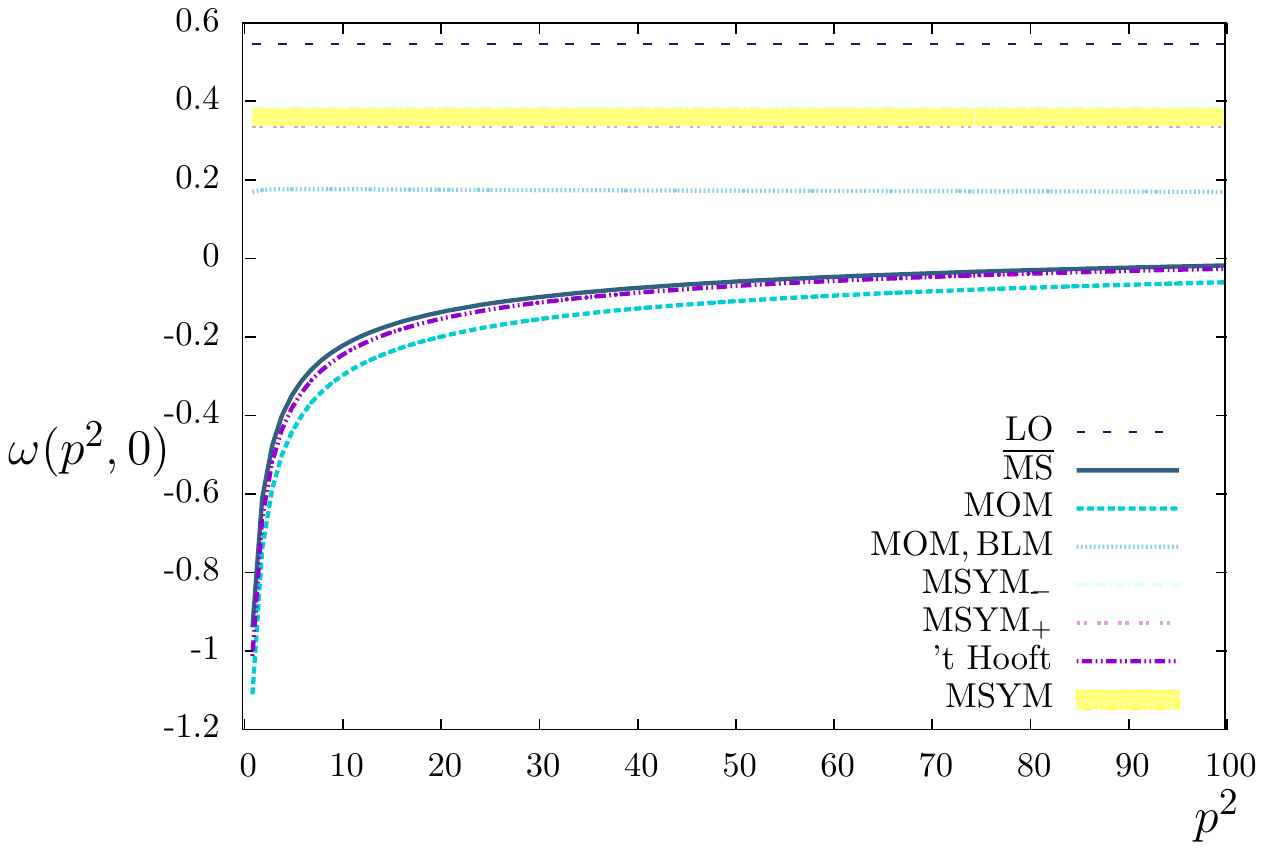}
\includegraphics[scale=0.8]{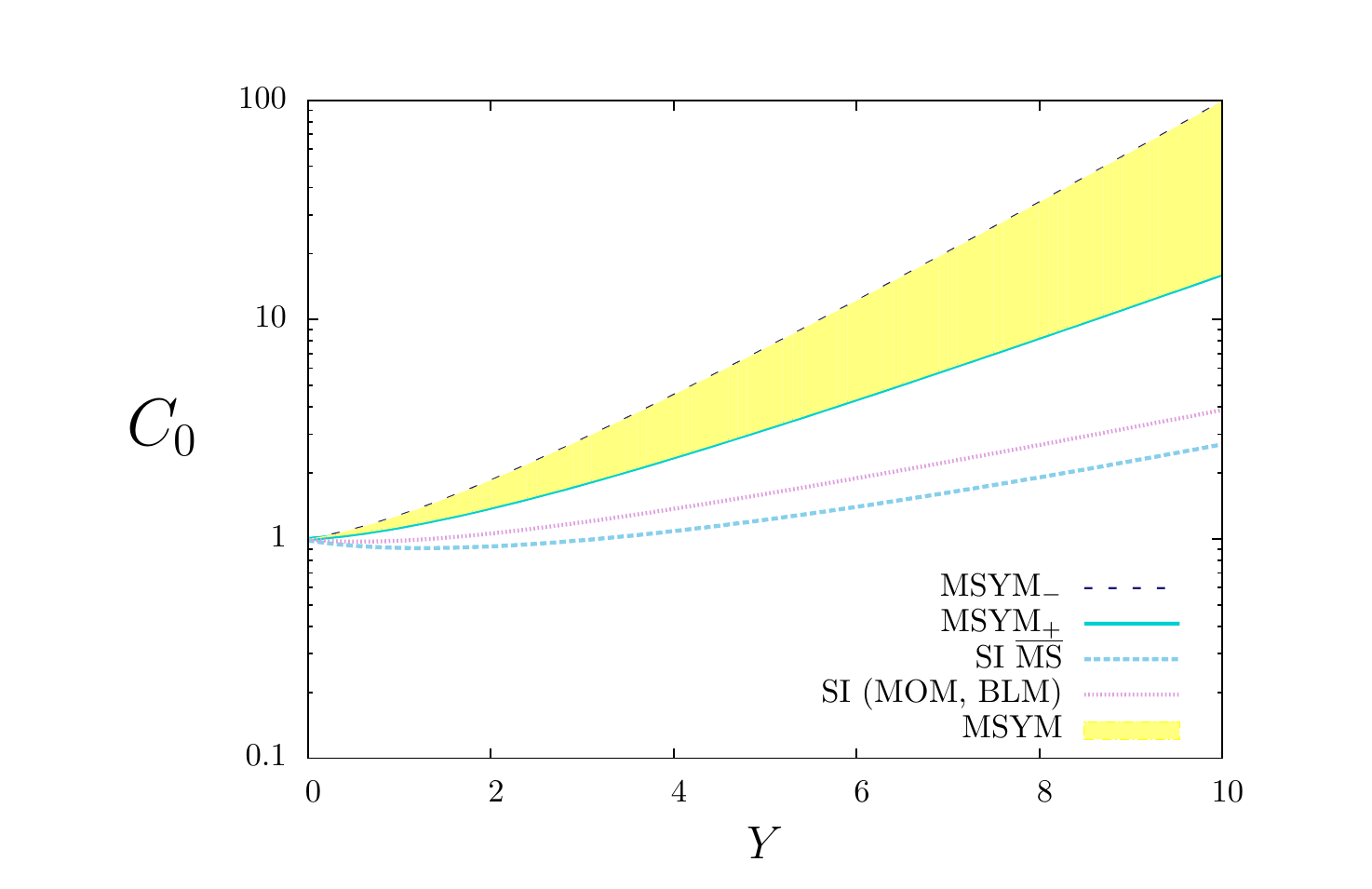}
\caption{\emph{Top:} Intercept vs jet resolution $\vec{p}^2$ for different renormalization schemes in QCD and MSYM; \emph{Bottom:} Growth of the cross section with rapidity separation $Y$.}
\label{1}
\end{figure}

\begin{figure}[h!]
\centerline{\includegraphics[scale=1]{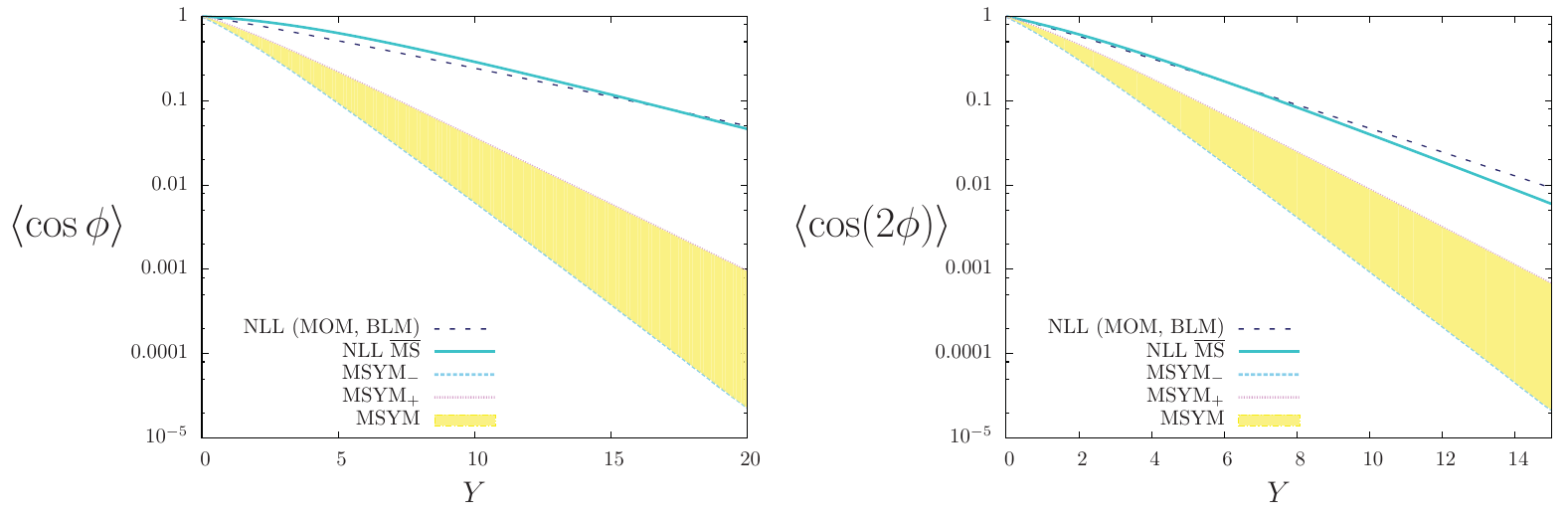}}
\caption{ Evolution of $\langle\cos \phi\rangle$ and $\langle\cos (2\phi)\rangle$ in QCD and MSYM.}
\label{2}
\end{figure}

\begin{figure}[h!]
\centerline{\includegraphics[scale=1.05]{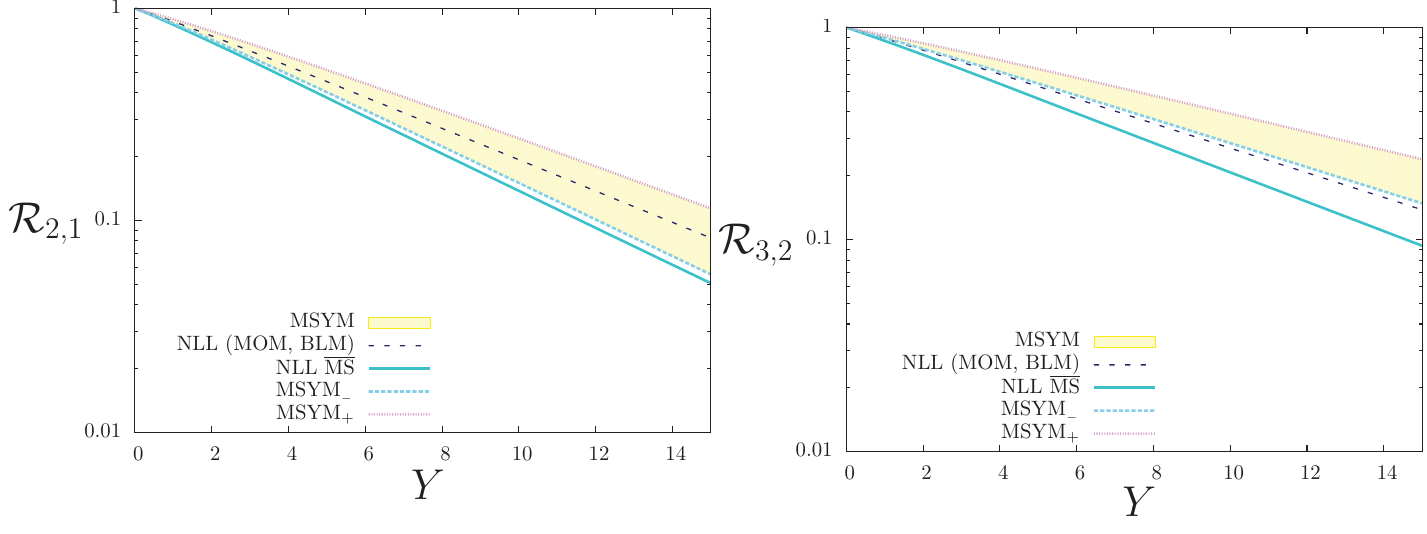}}
\caption{Evolution of ${\cal R}_{2,1}$ and ${\cal R}_{3,2}$ in QCD and MSYM.}
\label{3}
\end{figure}

In \textsc{Figs.} \ref{1}-\ref{3} we summarize the results obtained. ${\cal N}=4$ results appear as a yellow band because we take the coupling $\lambda$ to vary between $\lambda=\bar{\alpha}_s(\vec{q}^2/4)$ (MSYM$_-$) and $\bar{\alpha}_s (4\vec{q}^2)$ (MSYM$_+$). Results for QCD in the planar limit are also shown.\\

In \textsc{Fig.}~\ref{1} it is shown how BLM gives scale-invariant behaviour for the pomeron intercept $\omega(p^2,n=0)$ ---with a value of order 0.2---- and the fact that MSYM cross section grows faster than that of QCD for any renormalization scheme (RS) as a result of the dominance of real emissions and the higher multiplicity associated to extra supersymmetric fields. In \textsc{Fig.}~\ref{2} the lower order moments of the cross section in azimuthal angle are compared in QCD and ${\cal N}=4$. Tagged jets are less correlated in ${\cal N}=4$, again because of the higher final-state multiplicity. Due to the instability of the $n=0$ component, MOM-BLM results do not exhibit any clear effect with respect to other RS's. However, as shown in \textsc{Fig.}~\ref{3}, when plotting the ratios ${\cal R}_{n,m}$, from which reliable information can be obtained, BLM makes QCD results systematically closer to those of ${\cal N}=4$ SYM in all the range of rapidities, a feature also seen in a big number of low order ratios.

\section{Conclusions}
Study of dijet production at large rapidity separation through highly convergent ratios of azimuthal correlations in ${\cal N}=4$ SYM allows us to study the conformal dynamics of the Regge limit and analyze which renormalization schemes capture best the bulk of conformal contributions. In QCD, BLM procedure generates results very similar to those of ${\cal N}=4$ SYM ones in the studied kinematical window. This indicates that BLM is the natural renormalization prescription, at least for these observables, and brings hope that for well chosen quantities AdS/CFT results could be of some use to describe collider phenomenology. Other observables should be studied in future work to investigate this claim.

\bibliographystyle{normal}
\bibliography{Lowx2011_temp.bib}{}

\end{document}